\documentclass[aps,prd,twocolumn,superscriptaddress,amsfont,graphicx,nofootinbib,preprintnumbers]{revtex4}%
\usepackage{color,graphicx,epsfig}
\usepackage{ifpdf}
\usepackage{amsmath}
\usepackage{bm}
\usepackage{color}
\usepackage[english]{babel}
\usepackage{graphicx}%
\usepackage{amsfonts}%
\usepackage{amssymb}
\usepackage{braket}
\usepackage{hyperref}

\definecolor{nicered}{rgb}{0.7,0.1,0.1}
\definecolor{nicegreen}{rgb}{0.1,0.5,0.1}
\hypersetup{colorlinks,citecolor= nicegreen,linkcolor= nicered}

\newcommand{\beq}{\begin{equation}}
\newcommand{\eeq}{\end{equation}}
\newcommand{\bea}{\begin{eqnarray}}
\newcommand{\eea}{\end{eqnarray}}

\newcommand{\eref}[1]{Eq.~\eqref{eq:#1}}

\newcommand{\cref}[1]{Chapter~\ref{ch:.#1}}
\newcommand{\tref}[1]{Table~\ref{tab:#1}}

\newcommand{\nnl}{\nonumber \\}

\newcommand{\ba}{\begin{array}}  
\newcommand{\ea}{\end{array}} 
\newcommand{\be}{\begin{eqnarray}}  
\newcommand{\ee}{\end{eqnarray} }  
\newcommand{\bal}{\begin{align}}
\newcommand{\eal}{\end{align}}   
\newcommand{\bi}{\begin{itemize}}  
\newcommand{\ei}{\end{itemize}}  
\newcommand{\ben}{\begin{enumerate}}  
\newcommand{\een}{\end{enumerate}}  
\newcommand{\bc}{\begin{center}}
\newcommand{\ec}{\end{center}} 
\newcommand{\bt}{\begin{table}}
\newcommand{\et}{\end{table}}  
\newcommand{\btb}{\begin{tabular}}
\newcommand{\etb}{\end{tabular}}

\newcommand{\cO}{{\mathcal O}} 
 
\newcommand{\cL}{{\mathcal L}} 
 
\newcommand{\cM}{{\mathcal M}}

\newcommand{\gev}{\mathrm{GeV}}
\newcommand{\tev}{\mathrm{TeV}}

\def\pa{\partial}

\def\mysection#1{{{\bf #1}.~}}
\begin{document}

\def\LjubljanaFMF{Faculty of Mathematics and Physics, University of Ljubljana,
 Jadranska 19, 1000 Ljubljana, Slovenia }
\def\LjubljanaIJS{Jo\v zef Stefan Institute, Jamova 39, 1000 Ljubljana, Slovenia}
\def\CERN{CERN, Theory Division, CH-1211 Geneva 23, Switzerland}
\def\LPT{Laboratoire de Physique Th\'eorique, B\^at. 210, Universit\'e Paris-Sud, 91405 Orsay, France}

\title{Di-photon portal to warped gravity}

\author{Adam Falkowski}
\email[Electronic address:]{adam.falkowski@th.u-psud.fr}
\affiliation{\LPT}

\author{Jernej F.\ Kamenik}
\email[Electronic address:]{jernej.kamenik@cern.ch}
\affiliation{\LjubljanaIJS}
\affiliation{\LjubljanaFMF}


\date{\today}
\begin{abstract}
The diphoton excess around $m_X=750$ GeV observed by ATLAS and CMS can be interpreted as coming from a massive spin-2 excitation.
We explore this possibility in the context of warped five-dimensional  models with the Standard Model (SM) fields propagating in the bulk of the extra dimension.
The~750~GeV resonance is identified with the first Kaluza-Klein (KK) excitation of the five-dimensional graviton that  is parametrically lighter than KK resonances of SM fields.  
Our setup makes it possible to realize non-universal couplings of the spin-2 resonance to matter,  
and thus to explain non-observation of the 750 GeV resonance in leptonic channels. 
Phenomenological predictions  of the model depend on the localization of  fields in the extra dimension.
If, as required by naturalness arguments, the zero modes of the Higgs and top fields are localized near the IR brane, one expects large branching fractions  to $t \bar t$, $hh$, $W^+W^-$ and $ZZ$ final states.
Decays to $Z \gamma$ can also be observable when the KK graviton couplings to the SM gauge fields are non-universal. 
\end{abstract}

\maketitle

\mysection{Introduction}
\label{sec:intro}
The searches for diphoton resonances in the $\sqrt s=13$~TeV and 8~TeV LHC data have observed an excess around $m_{\gamma\gamma}\sim750$ GeV~\cite{ATLAS-CONF-2015-081,CMS:2015dxe,CMS:2016owr}, which can be interpreted in terms of a resonance ($X$) with~\cite{Franceschini:2015kwy, Falkowski:2015swt, Kamenik:2016tuv, others}
\beq
\label{eq:gammagamma}
\sigma(pp\to X)_{13\rm TeV}\times {{\rm Br}}(X\to\gamma\gamma)\approx 5 \ {\rm fb}\,.
\eeq
Among the possible spin assignments of $X$ allowed by Landau-Yang, the spin-zero case has so far received the most attention. On the other hand, perhaps the more intriguing possibility of a new massive spin-2 particle has so far been much less explored (see however~\cite{Han:2015cty,Giddings:2016sfr})\,, and is the main focus of the present work. Generically, massive spin-2 states are expected to appear as excited composite bound states of some strong dynamics, such as low energy QCD. Such interpretations face two severe challenges: (1) generically in strongly coupled scenarios one expects a plethora of other (lighter) states, which have so far not been observed; (2) the basic properties of $X$ such as its mass and couplings to SM fields cannot be expressed in terms of the (unknown) fundamental parameters and degrees of freedom of the theory.  

Fortunately, the dual picture of (approximately conformal) strongly coupled theories in terms of warped geometries in higher dimensions offers a tractable possibility to explore the phenomenology of massive spin-two resonances (now represented as KK excitations of the graviton) using perturbative methods. Here we follow this route and analyze the phenomenology of the lowest KK graviton excitation within warped extra-dimensional models in light of the LHC di-photon excess. In particular we want to address the following questions: (1) Under which conditions can the first KK graviton excitation be parametrically lighter than the rest of the KK spectrum and with the right properties to accommodate the di-photon excess; (2) What are the theoretical and phenomenological implications of KK graviton coupling non-universality as hinted at by null-results of the searches for 750~GeV resonances in other final states, especially in  di-leptons.   

In tackling these issues, we first briefly review the general spin-2 formalism and apply it to accommodate the LHC di-photon excess in light of other existing experimental constraints.  
We also discuss quantitatively the issue of perturbative unitarity loss in the presence of non-universal spin-2 couplings to matter.  
We then reconsider the model of Ref.~\cite{Giddings:2016sfr} with a warped extra dimension and with all the SM fields confined to the IR brane, which predicts universal couplings of KK gravitons to all SM matter fields. 
We discuss the tension due to experimental limits on the 750~GeV resonance from the di-lepton channel. 
We also point out that the hierarchy problem is not solved in this model. 
Addressing these short-comings requires some of the SM fields to be localized away from the IR brane,  and consequently SM gauge fields need to live in the bulk.
With this in mind, we propose a model with SM fields in the bulk supplemented by brane kinetic terms for gravity~\cite{Dvali:2000hr}, which allow for the first KK graviton excitation to be parametrically the lightest. We present a number of  phenomenologically viable parameter benchmark points and analyze their predictions in some detail. Finally, we summarize our main findings and conclude.

~

\mysection{Review of spin-2 formalism}
\label{sec:spin2}

A spin-2 particle can be represented by  a symmetric  tensor field, here  denoted as  $X_{\mu \nu}$.   
For a massive particle propagating in the flat space-time, the kinetic terms are described by the Fierz-Pauli Lagrangian: 
\bea 
\cL_{FP}  &= &
  {1 \over 2} (\partial_\rho X_{\mu \nu})^2  - {1 \over 2} (\partial_\rho X)^2 
-  (\partial_\rho X_{\mu \rho})^2  +   \pa_\mu X  \partial_\rho X_{\mu \rho}
\nnl  && 
-   {m_X^2 \over 2} (X_{\mu \nu})^2   +  {m_X^2 \over 2} X^2,  
\eea 
where $X = \eta_{\mu\nu} X_{\mu \nu}$, and $\eta_{\mu\nu} = {\rm diag}(1,-1,-1,-1)$ is the Minkowski metric  tensor.  
This form of the kinetic terms ensures that exactly 5 polarization states of the spin-2 particle  are propagating degrees of freedom. 
The interactions with the SM fields can be described by the following effective Lagrangian: 
\begin{align} 
\cL_{X} & =   
{c_V \over v } X_{\mu \nu} \left (  
{\eta_{\mu \nu} \over 4} V_{\rho \sigma}  V_{\rho \sigma}
-   V_{\mu \rho}  V_{\nu\rho}  \right ), 
\nnl 
 &\hskip-0.6cm -   {i c_f \over  2 v } X_{\mu \nu} \left (
 \bar f \gamma_\mu \overleftrightarrow D_\nu  f  
 -  \eta_{\mu \nu}   \bar f \gamma_\rho \overleftrightarrow D_\rho f   
  \right )
 \nnl   & \hskip-0.6cm +  
{c_H \over v } X_{\mu \nu} \left [ 
 2 D_\mu  H^\dagger D_\nu H  
 -  \eta_{\mu \nu} \left( D_\rho H^\dagger  D_\rho H -  V(H) \right) \right ]. 
\label{eq:spin2couplings}
\end{align}
Here $V \in G^a ,W^i, B$ denotes the SM $SU(3) \times SU(2) \times U(1)$ gauge fields,  $f \in (q_L,u_R, d_R,\ell_L,e_R)$ stands for the fermion fields, and $H$ is the Higgs doublet. 
The derivatives are covariant with respect to  the SM local symmetry and $\bar f \gamma_\mu \overleftrightarrow D_\nu f \equiv \bar f \gamma_\mu  D_\nu f - D_\nu \bar f \gamma_\mu   f$. 
The scale $v = 246$~GeV is inserted for dimensional reasons. 
For the massless graviton which mediates the  Einstein gravity, all the couplings $c_i$ are universal and equal to 
\beq
c_H = c_V = c_f = {v \over M_P} \approx 10^{-16},  
\eeq 
where $M_P = 2.44 \times 10^{18}$~GeV is the (reduced) Planck mass. 
However, for a massive graviton the couplings do not have to be universal in general; in this paper we will construct consistent  weakly coupled effective models where the couplings are non-universal.   

Given the couplings in Eq.~\eqref{eq:spin2couplings}, the partial decay widths of the spin-2 particle to SM mass eigenstates is given by  \cite{Han:1998sg,Lee:2013bua}: 
\bea 
\label{eq:gamma}
\Gamma(X \to h h) & = & 
{ c_H^2  m_X^3\over 960 \pi v^2} \left ( 1  - {4 r_{h}} \right )^{5/2} , 
\nnl 
\Gamma(X \to f \bar f )  &= & 
{ m_X^3 \over 320 \pi v^2} \left ( 1 -  {4r_{f}} \right )^{3/2}\left [    \right . \nnl && \left . 
 \left (c_{f_L}^2 + c_{f_R}^2 \right )  \left ( 1 -  {2  r_{f} \over 3 } \right ) 
 +   c_{f_L} c_{f_R} {20   r_{f} \over 3 }  \right ], 
\nnl 
\Gamma(X \to ZZ )   &= &   
{ m_X^3 \over 80 \pi v^2} \sqrt{1 - 4 {r_{Z}}} 
\left [ 
 c_{ZZ}^2 + {c_H^2 \over 12}  
   \right . \nnl && \left . 
 + {r_{Z} \over 3  }  \left ( 3 c_H^2 +  20 c_H c_{ZZ}  - 9 c_{ZZ}^2 \right )   
  \right . \nnl && \left .  
 + {2 r_{Z}^2 \over 3  }  \left ( 7 c_H^2 -  10  c_H c_{ZZ}  + 9 c_{ZZ}^2 \right ) 
 \right ] ,
 \nnl 
 \Gamma(X \to Z \gamma )   &= &   
  {c_{Z \gamma}^2 m_X^3 \over 40 \pi v^2} 
   \left ( 1  - {r_{Z}} \right )^{3} \left ( 1 + {r_{Z} \over 2 } + {r_{Z}^2 \over 6 } \right ) ,
 \nnl
\Gamma(X \to \gamma \gamma )   &= &   {c_{\gamma \gamma}^2 \over 8 c_G^2} \Gamma(X \to GG)
 =  {c_{\gamma \gamma}^2 m_X^3 \over 80 \pi v^2},
\eea 
where $r_{i} \equiv m_i^2 / m_X^2$\,.
Furthermore,  $\Gamma(X \to W^+ W^-) = 2 \Gamma(X \to Z Z)$ with $m_Z \to m_W$ and $c_{ZZ} \to c_W$.
The couplings to $Z$ and $\gamma$ can be expressed by the ones in the electroweak basis as 
$c_{\gamma \gamma} =  s_\theta^2 c_W + c_\theta^2 c_B $, 
$c_{Z Z} = c_\theta^2 c_W+  s_\theta^2 c_B$, 
$c_{Z \gamma} = c_\theta s_\theta (c_W - c_B)$, where $s_\theta$ and $c_\theta$ are the sine and cosine of the Weinberg angle. 
Note that the spin-2 decay to $Z \gamma$ occurs only for non-universal couplings.
Also note that decays to $hh$ always imply decays to $WW$ and $ZZ$ with at least a comparable branching fraction. 
This is because the spin-2 coupling to the Higgs boson is always accompanied by couplings to the Goldstone components of the Higgs doublet which, via the equivalence theorem, can be interpreted as  the longitudinal polarizations of W and Z.

The  production cross section of the spin-2 particle at the LHC can be written as 
\begin{align}
\sigma (p p \to X)_{E_{\rm LHC}}  &=
 {  \pi  m_X^2  \over v^2   E_{\rm LHC}^2}  \left [ 
  {1 \over 16} k_{GGX}  c_G^2  L_{GG} \left ( m_X^2 \over  E_{\rm LHC}^2 \right )  
  \right . \nnl  &+  \left .
   {1 \over 24} \sum_q k_{qqX}  (c_{q_L}^2 + c_{q_R}^2)  L_{q \bar q} \left ( m_X^2 \over  E_{\rm LHC}^2 \right )  
  \right ], \nnl
\end{align}
where the first term in the square bracket originates from gluon fusion, 
and the second from $q \bar q$ annihilation.  
Here, $ E_{\rm LHC} $ is the center-of-mass energy of proton-proton collisions at the LHC, $L_i$ are the parton luminosity functions, and $ k_{i}$ are the QCD k-factors which we take to be $ k_{GGX} \simeq k_{qqX} \approx 1.6$~\cite{Mathews:2005bw, Gao:2010bb}.  
For example,  using the NLO NNPDF2.3.2 pdf set~\cite{Ball:2014uwa} for gluon fusion production one finds  
$\sigma (p p \to X)_{13\rm TeV}  \approx  1.2  \times 10^{4} c_G^2$~pb.  
The numerical value of $c_G$ that reproduces the di-photon excess in Eq.~\eqref{eq:gammagamma} depends on ${\rm Br}(X \to \gamma \gamma)$, which is arbitrary at this point. We find
\beq
c_G \approx 3.1 \times 10^{-3} \sqrt{\frac{4.4 \times 10^{-2}}{{\rm Br}(X \to \gamma \gamma)}}\,.
\eeq
where we have chosen a reference value of ${\rm Br}(X \to \gamma \gamma)$ corresponding to the universal graviton coupling limit ($c_i = c_X$), c.f. Table~\ref{tab:RS_br}\,.
We see that requiring $c_G \lesssim 1$ implies ${\rm Br}(X \to \gamma \gamma) \gtrsim 10^{-7}$, otherwise the scale suppressing the spin-2 couplings to gluons would be  below the electroweak scale.  The opposite limit ${\rm Br}(X \to \gamma \gamma) \lesssim 1$ implies that $c_G \gtrsim 6.6 \times 10^{-4}$ corresponding to a suppression scale below $\sim 400$~TeV.

Spin-2 particles produced at the LHC are polarized. 
If the polarization is measured along the beam axis, at LO in QCD
only $ h = \pm 2$ helicities are produced in gluon fusion, 
and only $ h = \pm 1$ helicities are produced in $q \bar q$ annihilation.   
Polarization determines the angular distribution of the photons to which the spin-2 particle decays. 
We define  $\theta^*$ as the angle between the decay direction and the beam axis in the rest frame of the decaying particle.
The  $\theta^*$ distributions  are then given by \cite{Artoisenet:2013puc}: 
 
 \be 
 \label{eq:theta}
 {d \sigma_{GG\to X \to \gamma \gamma} \over d \cos \theta^*}  &\sim  & 1 + 6 \cos^2\theta^* + \cos^4 \theta^*, 
 \nnl 
 {d \sigma_{q \bar q \to X \to \gamma \gamma} \over d \cos \theta^*}  &\sim  & 1 -  \cos^4 \theta^*. 
   \eea
In gluon fusion production, the photons are strongly peaked in the forward directions, 
while this effect is absent for $q \bar q$ annihilation.

A field theory with an interacting spin-2 particle is always an effective theory with a limited range of validity. 
That is because tree-level amplitudes involving the spin-2 particle grow with energy, the couplings in \eref{spin2couplings} having canonical dimensions 5. 
For universal couplings, $c_i = c_X$ the maximum cut-off scale $\Lambda_{\rm cut}$ of the effective theory is of the order $\Lambda_{\rm cut} \sim {4 \pi  v / c_X}$, 
at which scale amplitudes cease to be perturbatively unitary. 
Interestingly, Ref.~\cite{Artoisenet:2013puc} pointed out that for non-universal couplings of the graviton one observes a worse high energy behavior of the amplitudes. 
In particular, amplitudes of spin-2 particle production may grow as fast as $\cM \sim c_i E^3/m_X^2 v$, leading to a faster unitarity loss.   
One example is the process $q_L \bar q_L \to X G$ with massless left-handed quarks in the initial state. 
 The s-wave amplitude for producing a helicity-0 graviton in this process is given by 
\beq
 \cM_0 (0,\pm 1)= 
\pm  g_s \left (c_G - c_{q_L} \right ) {s^{3/2} \over 32 \sqrt 3  \pi m_X^2  v} . 
\eeq
where $s$ is the invariant mass of the $q \bar q$ pair. 
As a consequence,  perturbative unitarity is lost (${\rm Re}[\mathcal M_0(0,\pm 1)]>1/2$) at 
\beq
\Lambda_{\rm cut}  \approx \left ({16 \sqrt 3 \pi v m_X^2 \over g_s |c_G - c_{q_L} |} \right )^{1/3}
\approx 3 m_X \left ( v \over m_X |c_G - c_{q_L}| \right )^{1/3}. 
\eeq 
For example, for  $m_X = 750$~GeV,  $c_{q_L} \approx 0$ and $v/|c_G| \approx 10$~TeV, one needs a  cut-off below  $\Lambda \sim 8$~TeV, and not at  $4 \pi v/|c_G| \sim 100$~TeV as one may naively expect. 
These  considerations motivate studies of more fundamental scenarios from which massive spin-2 states can emerge.
One such construction is the Randall-Sundrum  (RS) model discussed in the remainder of this paper. 

~

\mysection{RS with SM on IR brane}
\label{sec:RS-IR}

The RS model  \cite{Randall:1999ee} is a gravity theory formulated in the 5D spacetime. 
The fifth dimension is  an interval, $y \in [0,L]$, where the boundary at $y=0(L)$ is referred to as the UV(IR) brane. 
The 5D Lagrangian is given by 
\begin{align} 
\label{eq:RS_l}
\cL_{\rm RS} & =    \sqrt{g}  M_*^3\left [ 
 - {1 \over 2} R_5  + 6 k^2  \right ]  
\nnl  &+
M_*^3 \sqrt{-g_4} \left [ \delta(y-L) - \delta(y) \right ]  \left [ 3 k   - D^\alpha n_\alpha \right ]  . 
 \end{align}
 The gravitational  degrees of freedom are described by the 5D metric field $g$. 
$R_5$ is the 5D Ricci scalar constructed from that metric, $M_*$ is the 5D Planck mass, 
and the scale $k$ sets the magnitude of the (negative) 5D cosmological constant.    
 Finally,  $g_4$ is the 4D metric field obtained  by projecting  $g$ on the brane. 
We also included the Gibbons-Hawking terms  $D^\alpha n_\alpha$, where $n_\alpha =  g_{55}^{-1/2}(0,0,0,0,1)$, 
 which is necessary to arrive at consistent Einstein equations on a manifold with boundaries \cite{Gibbons:1976ue}.
This Lagrangian leads to the Einstein equations for  $g$ whose solution is  a slice of AdS${}_5$ metric.   
We parametrize the 5D metric as 
\beq
\label{eq:metric}
ds^2 = a^2(y) \left ( \eta_{\mu \nu} + h_{\mu\nu}(x,y) \right )  dx_\mu   dx_\nu  
-  dy^2 .
\eeq 
Here  $a(y) = e^{- k y}$ is called the warp factor, and $h_{\mu\nu}(x,y)$ describes perturbations of the metric around the AdS${}_5$ background.\footnote{In general one should also include a scalar degree of freedom, the so-called radion, associated with the overall length of the 5th dimension.
In the current setup the radion is massless. 
In the following we assume it obtains a large enough mass by some mechanism (see e.g. \cite{Goldberger:1999uk}) that does not affect significantly the background solution and does not play any role in the  phenomenology of the 750~GeV resonance.}

In the original RS construction the SM fields are assumed to be confined to the IR brane: 
$\cL \supset  M_* \sqrt{-g_4} \cL_{\rm SM}  \delta (y-L)$ where $\cL_{\rm SM}$ is the usual SM Lagrangian. 
To derive phenomenological predictions of this model, one expands the metric perturbations into a discrete set of KK modes:  $h_{\mu \nu}(x,y)  = \sum_{n=0}^\infty X_{\mu \nu}^{(n)}(x) f_n(y)$, where the KK profiles $f_n$ satisfy the equation 
\beq
\label{eq:graviton_eom}
\partial_y^2  f_n  + 4  {a' \over a}   \partial_y f _n  + m_n^2 a^{-2} f_n =  0, 
\eeq 
together with the boundary conditions $\pa_y f_n(0) = \pa_y f_n(L) =0$, 
and the orthonormality condition  
\beq
\label{eq:normIR}
M_*^3 \int_0^L dy  a^2(y) f_n(y) f_m(y)   = 4  \delta_{nm}\,. 
\eeq
This way, each $X_{\mu \nu}^{(n)}$ is a canonically normalized spin-2 field fitting in the general formalism described in the previous section.
 \eref{graviton_eom} admits the zero-mode solution with $m_n =0$, and a flat profile 
$f_0 = [2 M_*^3(1- a_L^2)/k]^{-1/2}$, $a_L \equiv e^{-k L}$.  
This corresponds to the massless graviton mediating the Einstein gravity in 4D. 
For $m_n >0$ the solution satisfying $\pa_y f_n(0)  = 0$ can be written in terms of the Bessel functions, 
\begin{align}
\label{eq:graviton_profile}
f_n(y) &= A_n a^{-2}(y) \left [ 
Y_{1}\left (m_n \over k \right )  J_{2}\left (m_n \over a(y) k \right ) \right. \nnl
& \left. - J_{1}\left (m_n \over k \right ) Y_{2}\left (m_n \over a(y) k \right ) \right ] . 
\end{align}
Then the other  boundary condition  $\pa_y f_n(L)  = 0$  corresponds to the KK mass quantization condition which  can be written as 
\beq
Y_{1}\left (m_n \over k \right )  J_{1}\left (m_n \over a_L k \right )
- J_{1}\left (m_n \over k \right ) Y_{1}\left (m_n \over a_L k \right )  = 0\, . 
\eeq 
The quantization condition is solved by a discrete set of $m_n$ starting parametrically at  $\cO(k a_L)$. 
Numerically, one finds that  the first KK mode occurs at $m_1 \approx 3.8 k a_L$, 
and them $m_2 \approx 1.8 m_1$,  $m_3 \approx 2.7 m_1$, \dots.
The constant $A_n$ in \eref{graviton_profile} is fixed by the normalization condition in Eq.~\eqref{eq:normIR}. 
Given the KK profile $f_n(y)$,  the coupling of the graviton $n$-th mode to the matter on the IR brane  is given by 
\beq 
\label{eq:IR_cxn}
c_{X_n} =  {f_n (L) \over f_0} {v \over M_P} \,, \qquad M_P^2 = {M_*^3(1 + a_L^2)  \over 2 k}, 
\eeq 
where $M_P$ is the reduced Planck mass that sets the coupling strength of the massless graviton. 
For the first mode one finds to a good approximation 
\beq
\label{eq:IR_cx1}
c_{X_1}  \approx - {v \over a_L M_P} .  
\eeq 
The coupling is enhanced by $a_L^{-1}$ compared to the zero mode one 
because, in the RS scenario with a large warp factor, the lowest KK modes are sharply localized near the IR brane.     
Thanks to this enhancement,  the RS scenario with a large warp factor at the IR brane can address the 750~GeV  excess, with the diphoton resonance  identified as the first KK mode of the graviton, as previously discussed in Ref.~\cite{Giddings:2016sfr}.  
This model is extremely predictive, with basically no free parameters. 
Since all the graviton couplings are universal, the branching fractions are completely fixed; 
in particular, ${\rm Br}(X \to \gamma \gamma) \approx  4.4\%$. 
Then the IR warp factor $a_L$ is fixed to fit the observed production cross section of the resonance, c.f. \eref{IR_cx1},  the curvature scale $k$ is fixed to fit the 750~GeV resonance mass using  $m_{X_1} \approx 3.8 k a_L$, and the 5D Planck mass is fixed to fit the 4D Planck mass, c.f.  \eref{IR_cxn}. 
A set of parameters that nicely fits the ATLAS and CMS observations is the following 
\beq
\label{eq:IR_bench} 
a_L = 3.4 \times 10^{-14}\,, ~ 
k = 5.8 \times 10^{15}~\gev\,, \nonumber
\eeq
\beq
M_* = 4.1 \times 10^{17}~\gev .  
\eeq 
We refer to this benchmark point as {\bf IR}, to distinguish it from other RS benchmarks studied in the next sections. 
The resulting value of the universal graviton coupling to matter is $c_X = -0.003$, which implies   $\sigma(pp \to X)_{13\rm TeV} \approx 0.12\,$pb, 
and the diphoton rate  $\sigma(pp \to X \to \gamma \gamma)_{13\rm TeV} \approx 5.2\,$fb.
 The width of the resonance well below the experimental resolution: $\Gamma_X \approx 6\,$MeV. 
By choosing a larger (smaller) warp factor we can make the cross section and width smaller (larger), 
but the parameters ballpark  has to stay the same to match the observations. 
The KK graviton branching fractions are summarized in  the first column of~\tref{RS_br}. 

All in all, the RS model with the SM on the IR brane offers an interesting and very predictive spin-2 model for the 750~GeV resonance.  
There are however two issues with this scenario: one phenomenological, and one theoretical.
The phenomenological one is the tension with the dilepton resonance searches at 8~TeV and 13~TeV LHC. 
As we can see in \tref{RS_br}, this scenario sharply predicts 
$ \Gamma(X \to e^+e^- + \mu^+\mu^-)/\Gamma(X \to \gamma \gamma) = 1$, and thus $\sigma(pp\to X)_{8\rm TeV} \times {\rm Br}(X\to e^+e^- + \mu^+\mu^-)\approx 1.2$~fb. 
This is in tension with the ATLAS search~\cite{Aad:2014cka} for di-lepton resonances in LHC run-1 which found 
$\sigma(pp \to Z' \to e^+e^- + \mu^+\mu^-)_{8\rm TeV} \leq 1.3$~fb at 95\% CL. More recently ATLAS performed a similar search using the first 13TeV data, putting a bound on  $\sigma(pp \to Z' \to e^+e^- + \mu^+\mu^-)_{13\rm TeV} \leq 5.5$~fb at 95\% CL.~\cite{dileptons}.
Not discovering a 750~GeV  dilepton resonance in 2016 run-2 data  will ultimately falsify this model.   
The theoretical issue is that, for the parameters in~\eref{IR_bench}, the scale $\Lambda_X$ suppressing graviton couplings to matter and itself is very large, $\Lambda_X \approx 100$~TeV. 
Therefore, the cut-off scale at the IR brane (the energy scale up to which SM scattering amplitudes are perturbative)   is rather high, $\Lambda_{\rm cut} \sim 1000$~TeV. 
Therefore, the quantum corrections to the mass term of the Higgs field are cut off at a high $\Lambda_{\rm cut}$ scale, and the little hierarchy problem  of the SM, which was the original motivation of Ref.~\cite{Randall:1999ee},  is not addressed. 
In the next section we discuss a modification of the RS scenario that potentially addresses both of these issues.  

\begin{table}[h]
\bc 
\begin{tabular}{|c|ccccc|}
\hline
& \ IR \ & MIN & MED & MAX  & GMAX
\\  \hline\hline
$\gamma \gamma$  & 4.3 &  8.5 &  7.0  & 0.5 & 2.3 
\\  
$ZZ $ & 4.8 & 7.9 &  7.8 &  2.9   & 12 
\\  
$WW$ &9.5 &  16 &  15   & 5.6 & 21  
\\  
$Z\gamma $ & 0 & 0 &  0 &  0 & 1.1 
\\  
$hh$ & 0.3 & 0 &  0.4   & 1.4 &6.9 
\\  
$tt$ &5.1 & 0 &  8.3  & 85 & 56 
\\  
$bb$ &6.4  & 0 &  5.2      & 0.4  & 0.04 
\\  
$jj $ &66 & 68 & 61    & 4.5  & 0.5 
\\  
$e^+e^-+\mu^+\mu^-$ & 4.3 & 0 &  0  & 0 & 0 
\\  \hline
\end{tabular}
\ec
\caption{
\label{tab:RS_br}
Branching fractions (in percent) of the first graviton KK mode to various SM final states for the IR, MIN, MED, MAX, and GMAX benchmarks described in the text.   
}
\end{table} 

~

\mysection{RS with SM in bulk}
\label{sec:RS-bulk}

The prescription to reduce the KK graviton couplings to leptons  is to localize the two in different  points in the extra dimension.
This will be the case when leptons are localized away from the IR brane. 
One possibility is to localize the entire SM at the UV brane, but then the couplings of the KK gravitons to matter are extremely suppressed and have no phenomenological relevance. 
A more fruitful direction is to promote  the SM gauge and matter fields to 5D fields which 
propagate in the bulk of the extra dimension \cite{Davoudiasl:1999tf, Pomarol:1999ad, Bajc:1999mh,Chang:1999nh}, as is already the case for gravity.  
The SM particles are then identified with the zero modes of the 5D fields, while their KK modes must be heavy enough to have avoided detection so far. 
The setup contains free parameters (5D mass terms)  that allow one to shape the zero mode profiles of fermion and scalar fields (on the other hand, the zero mode profiles of unbroken gauge fields are always flat in the 5th dimension).  
This framework has been extensively studied in the past; in fact much more than the original RS  model with the SM localized on the IR brane. 
Historically, the main motivation was the fact that it allows one to address the SM fermion mass hierarchies, by controlling the overlap of the fermion zero mode profiles with the IR brane where the Higgs field is assumed to reside~\cite{Grossman:1999ra}.  
Moreover, it also allows one to address the hierarchy problem in the so-called gauge-Higgs unification scenario, when the Higgs boson arises from the 5th component of a gauge field from an extended gauge group~\cite{Manton:1979kb}.    

The obvious problem with the idea in the context of the spin-2 explanation of the 750 GeV excess is that the minimal scenario predicts  the first KK modes of the SM gauge fields to be {\em lighter} than first graviton KK mode. 
This is clearly unacceptable phenomenologically. 
However, it is possible to make the first graviton KK mode to be parametrically lighter than the gauge KK modes. 
Notice that the symmetries of the RS framework allow one to introduce brane kinetic terms for the graviton \cite{Dvali:2000hr,Kiritsis:2002ca,Davoudiasl:2003zt}. 
We add to the RS Lagrangian in~\eref{RS_l} the following terms  
\beq 
 \Delta \cL_{\rm RS}= 
  - {1 \over 2} M_*^3  \sqrt{-g_4}  R_4  \left [  r_0 \delta (y) + r_L \delta (y-L) \right ], 
\eeq 
where $R_4$ is the 4D Ricci scalar constructed out of the 4D metric $g_4$ induced at the branes, and $r_0$, $r_L$ are parameters of dimension ${\rm mass}^{-1}$.  
The presence of brane kinetic terms does not affect the equation of motion~\eref{graviton_eom}  for the KK profile of the graviton.  
What changes are the boundary conditions, which now read 
$\partial_y f_n(0)  = - r_0 m_n^2 f_n(0)$, 
$a_L^3  \partial_y f_n(L)  =  r_L m_n^2 f_n(L)$, 
and the orthonormality conditions, 
which read
\beq 
\label{eq:normBulk}
 M_*^3  \int_0^L dy a^2 f_n f_m    \left ( 1 +  r_0 \delta (y) +   r_L \delta (y-L) \right )   = 4 \delta_{nm}\,. 
\eeq
The zero-mode solution with $m_n=0$ and a constant profile  is retained in the spectrum, however its normalization is changed: 
\beq 
f_0 =   {2 \over M_P}, \qquad M_P^2 =  M_*^3\left ( 
{1 - a_L^2 \over 2 k}  + r_0  + a_L^2 r_L \right ).  
 \eeq 
Note that the relation between the observable 4D Planck mass $M_P$ and the parameters in the 5D Lagrangian is also affected. 
The KK profiles for the massive modes which  satisfy the boundary condition at $y=0$ can be written as 
\bea 
f_n(y)  &= & A_n a^{-2} \left [ 
\left ( Y_{1}\left (m_n \over k \right )  + m_n r_0 Y_{2}\left (m_n \over k \right ) \right )J_{2}\left (m_n \over a k \right )
\right . \nnl && \left . 
- \left ( J_{1}\left (m_n \over k \right )  + m_n r_0 J_{2}\left (m_n \over k \right ) \right ) Y_{2}\left (m_n \over a k \right ) \right ]. 
\eea
The boundary condition at $y=L$ then determines the quantization condition 
{
\begin{align}
\label{eq:RSBKT_quantization}
&\left [ Y_{1}\left (m_n \over k \right )  + m_n r_0 Y_{2}\left (m_n \over k \right ) \right ] \nnl
&\times \left [ J_{1}\left (m_n \over a_L k \right ) - {r_L m_n \over a_L} J_{2}\left (m_n \over a_L k \right ) \right ]  \nnl
& = 
\left [ J_{1}\left (m_n \over k \right )  + m_n r_0 J_{2}\left (m_n \over k \right ) \right ] \nnl
&\times \left [ Y_{1}\left (m_n \over a_L k \right ) - {r_L m_n \over  a_L} Y_{2}\left (m_n \over a_L k \right ) \right ] , 
\end{align}
}
while  the normalization constant $A_n$ is determined by the orthonormality condition in Eq.~\eqref{eq:normBulk}. 
When $r_0$ and $r_L$ are small in units of $1/k$, the graviton spectrum is only slightly modified compared to the original RS. 
The interesting things happen when both brane kinetic terms are large, $k r_{0,L} \gg 1$.\footnote{%
One should note that for a large IR brane kinetic term, $k r_L >1/2$,  the radion becomes a ghost in this setup, in the sense that its kinetic terms are not positive definite \cite{Csaki:2000zn,Luty:2003vm,George:2011sw}. See Appendix for more details.  
We will assume that there exists a stabilization mechanism that gives radion positive-definite kinetic terms for $k r_L  \gg 1$, although that has not been demonstrated in the literature so far.
We thank Kaustubh Agashe and Riccardo Rattazzi for pointing out the radion ghost problem to us.
 }
One can then show that the first massive solution of the quantization condition in \eref{RSBKT_quantization} is approximately given by 
\beq 
\label{eq:gravitonmass}
m_{X_1} \approx  {2 a_L k \over  \sqrt{k r_L}}    . 
\eeq 
Thus, in the RS scenario with large brane kinetic terms the mass of the  first graviton KK mode  
is parametrically suppressed compared to the rest of the KK tower (which starts at $\cO(k a_L)$). 
If the bulk SM fields do not have large brane kinetic terms, the mass of the first graviton KK mode will be suppressed compared to all other KK states in the theory.  
This mechanism can thus explain why the spin-2 KK  state is the first one to be discovered at the LHC. 
For example, the first KK mode of the SM gauge field is predicted at $m_{V_1} \approx 2.4 k a_L$. 
Then, choosing $k r_L \sim 10$ is enough to push the first gauge KK mode up to  $m_{V_1} \sim 3$~TeV. 
Note that a mild hierarchy of parameters is always present  in the RS scenario: to obtain the large warp factor at the IR brane, $a_L \sim 10^{-15}$, one needs to choose the length of the extra dimension $L$ to be large in units of inverse curvature, $k L \sim 35$. 
We argue that similarly mild hierarchies for graviton brane kinetic terms are enough to render phenomenologically viable the RS model with bulk SM fields and a 750 GeV KK graviton.  

For large $r_{0,L}$, the normalized profile of the first graviton KK mode  can be approximated as 
\beq
\label{eq:gravitonprofile}
f_1(y) \approx     {2 \over M_P}    a_L  
\left ( (1 - a^{-4}) a_L^2 \sqrt{r_0 \over r_L} + (1 -  a_L^4 a^{-4} ) \sqrt{r_L \over r_0}  \right ).
\eeq
Much like in the original RS scenario, it is strongly peaked ($\sim \exp{4 k y}$) towards the IR brane, and strongly suppressed near the UV brane. 
Therefore, by localizing some SM fields away from the IR brane we can suppress their couplings. 
Consider the normalized zero mode profile of the form: 
\beq
\label{eq:SMprofile}
f_{{\rm SM}, \alpha} (y) =  \sqrt{2 k \alpha \over 1 - a_L^{2\alpha}}  e^{ - \alpha k y}, 
 \eeq 
 which is normalized as  $\int_0^L dyf_{{\rm SM}, \alpha} (y) ^2 = 1$.  
For SM matter fields the parameter $\alpha$ controlling the localization is arbitrary, as it is  fixed by free parameters in the 5D Lagrangian.   
For  fermion fields, $\alpha$ is determined by the bulk mass term $M$ of the 5D fermion field: $\alpha = \pm M/k  + 1/2$ for a left/right-handed zero mode of  a SM fermion. 
Similarly, for scalar fields  the bulk mass term can be used to control localization of the zero mode,  though in this case the existence of the zero mode requires adding boundary mass terms with a fine-tuned magnitude.  
On the other hand, for unbroken gauge fields the zero mode profile is always flat, which corresponds to taking the limit  $\alpha \to 0$ in \eref{SMprofile}.
Using the normalization in \eref{spin2couplings}, 
the coupling of a fermionic zero-mode  to the first graviton KK mode is given by 
\begin{align}
\label{eq:calpha}
c_{\alpha} &=   {v \over M_P} \int_0^L f_{{\rm SM}, \alpha} (y) {f_1(y) \over f_0}
\nnl &\hskip-0.4cm \approx   - {v \over a_L M_P} 
\left [ \sqrt{r_0 \over r_L}  {\alpha \over  (\alpha -2)(1 - a_L^{-2\alpha})} 
 + \sqrt{r_L \over r_0}  { a_L^2 \over a_L^{2\alpha} -1 } \right ] .
\nnl 
\end{align}  
The first term  in the square bracket is relevant for $\alpha < 1$, and the second for $\alpha >1$. 
For sharp IR localization, $\alpha \ll 0$,  one finds  
$c_{\alpha} \approx   - ({v / a_L M_P})  \sqrt{r_0 / r_L}  {\alpha /  (\alpha -2)}$. 
In this case the scale controlling the  coupling to the first KK graviton is  $a_L M_P$. 
This is similar, up to a numerical factor depending on $\alpha$, $r_0$, $r_L$,  to the previously discussed  case of KK gravitons coupling to IR-localized matter fields in the original RS model. 
In the limit $\alpha = 0$ one finds     $c_{\alpha} \approx   - {(v / a_L M_P)}  \sqrt{r_0 / r_L} {(1 /  4 k L)}$. Thus, zero mode gauge fields, as well as matter fields with the flat profile, have a coupling to graviton suppressed by the factor $4 k L$ compared to IR-localized fields. 
For $0 < \alpha < 1$, when the profile is tilted towards the UV brane,  the coupling becomes suppressed by powers of the warp factor,  $c_{\alpha} \approx   - {(v a_L^{2\alpha}/ a_L M_P)}  \sqrt{r_0 / r_L}  {\alpha / (2- \alpha)}$. 
This switches to a universal behavior for  $\alpha > 1$:  $c_{\alpha} \approx   {(v a_L /  M_P)} \sqrt{r_L / r_0}$.  
In general, the zero modes with $\alpha > 0$ couple very weakly to the KK  gravitons,  and they  do not play any role in production and decay  of the 750 GeV resonance.    

We are ready to discuss concrete parameters choices that allow one to explain the 750 GeV excess in the RS model with SM gauge fields propagating in five dimensions. 
The diphoton resonance is identified with the first KK mode of the graviton with the KK profile in \eref{gravitonprofile} and the mass  in \eref{gravitonmass}.
Its couplings to the SM  are determined by  \eref{calpha} and depend  on the parameters $\alpha_i$ (specific for each SM particle) controlling the shape of the zero mode profile.  
The SM gauge fields:  gluons, W and Z bosons, and photons have a flat profile in the extra dimension corresponding to $\alpha_V =0$.  
To avoid constraints from dilepton resonance searches, we assume that  the lepton fields are localized near the UV brane, $\alpha_\ell > 0$, in which case they are practically decoupled from the graviton KK modes. 
In the best of all worlds,  the Higgs field should be localized near the IR boundary, 
so that the hierarchy problem can be addressed by the large warp factor. 
Then also the top quarks  should be IR localized, so that the top Yukawa coupling remains in the perturbative regime. 
This is however non-trivial to achieve. 
The reason is that the graviton couplings to gauge zero modes carry a suppression  factor  $4 k L \sim 130$, while the couplings to IR localized fields do not have this suppression. 
If the Higgs field was strictly IR-localized, the branching fraction of the 750 GeV resonance would be dominated by  decays to pairs of Higgs bosons and longitudinally polarized W and Z bosons. 
This would violate the LHC run-1 constraints from di-Higgs and di-boson resonance searches~\cite{Aad:2015kna, Khachatryan:2015cwa, ATLAS:2014rxa} (c.f.~\cite{Franceschini:2015kwy}).  
 As a consequence,  in this setup, only mild IR localization of the Higgs and the top is possible.  

\begin{table}
\bc
\begin{tabular}{|c|ccc|}
\hline
  & MIN & MED & MAX  
  \\ \hline \hline
$r_0$[1/k] & $100 $ & $120$ & $1700$ 
\\ 
$M_*$[GeV] & $4.1 \times 10^{17}$ & $3.9 \times 10^{17}$ & $1.6 \times 10^{17}$  
\\ \hline 
$\alpha_{t_R}$ & $\infty $  & $0$ & $-0.3$  
\\ 
$\alpha_{Q_L^3}$ & $\infty$  & $0$ & $0$   
\\ 
$\alpha_{H}$ & $\infty$  & $0$ & $-0.1$ 
\\ \hline 
$-c_G$ & $2.3 \times 10^{-3}$ &  $2.5 \times 10^{-3}$ &  $9.6 \times 10^{-3}$ 
\\ 
$\sigma (pp \to X )$[pb] & 0.06 & 0.08 & 1.1
\\ 
$\sigma (pp \to X \to \gamma \gamma)$[fb] & 5.3  & 5.3 & 5.4
\\ 
$\Gamma_X$[GeV] &$2 \times 10^{-3}$ & $3 \times 10^{-3}$ &  0.5 
\\ \hline 
\end{tabular}
\ec
\caption{
\label{tab:benchmarks}
The parameters for various benchmark models fitting the 750 GeV excess with the first KK graviton in the RS scenario. 
All benchmarks have $a_L = 10^{-15}$,  
$k=1.2 \times 10^{18}$~GeV, $r_L =10/k$. 
The lowest graviton KK modes are at  $m_1=m_{X} = 750$~GeV,   $m_{2} \approx 6$~TeV,
and lowest hypercharge KK mode is at $m_{V_1} \approx 2.9$~TeV. 
The SM fermion fields other then the right-handed top and and the 3rd generation quark doublet are assumed to be sharply localized at the UV brane. 
 }
\end{table}

We propose  3 concrete benchmarks: MIN, MED, and MAX, which represent various degrees of compromise between the naturalness and phenomenological requirements. 
The parameters of the 5D Lagrangian  are summarized in \tref{benchmarks},
and they are tuned to yield $m_1 = m_{X} = 750$~GeV and $\sigma(pp \to X)_{13 \rm TeV} {\rm Br}(X \to \gamma \gamma) \sim 5$~fb. 
The latter number can be easily adjusted up or down; for example, increasing (decreasing) the UV brane kinetic terms $r_0$ increases (decreases) the diphoton signal, without affecting other predictions significantly. 
The width of the 750 GeV resonance is always smaller than the experimental  resolution at the LHC for all benchmarks.  
All the benchmarks satisfy phenomenological bounds from LHC run-1 resonance searches in other decay channels, as summarized in \tref{RSBKT_rgamma}.  
The first  KK modes other than the graviton one are predicted to be at the scale $m_{V_1} \approx $~2.8 TeV, and correspond to KK excitations of the SM gauge fields. 
That mass scale can be raised by increasing the value of the IR brane kinetic term $r_L$.  
The predicted $X$ decay branching fractions are shown in the middle columns of \tref{RS_br}.

In the {\bf MIN} benchmark, the SM fermion and Higgs fields are all localized on the UV brane, which implies that  the hierarchy problem is not addressed at all. 
Only  the SM gauge fields have non-negligible couplings to the KK gravitons, 
and these couplings are universal. 
The sharp prediction of this scenario is ${\rm Br}(X \to \gamma \gamma) \approx  {\rm Br}(X \to ZZ)  \approx {\rm Br}(X \to WW)/2 \approx {\rm Br}(X \to GG)/8 \approx 0.08$, and no other decays of the 750 GeV particle are present. 
In the {\bf MED} benchmark,  the Higgs, the right-handed top, and the 3rd generation quark doublet all have a flat profile, much like the gauge fields, while the remaining fermions are localized at the UV brane. 
The phenomenological predictions turn out to be very similar as for the MIN benchmark. 
 The difference is that decays of the 750 GeV resonance to top and bottom quarks and to the Higgs boson are present.
However, with a flat profile of these SM fields, the branching fractions are probably too small for the top, bottom and Higgs decays to  ever be observable. 
That changes spectacularly for the {\bf MAX} benchmark, where the Higgs and the right-handed top are localized toward the IR brane (while the 3rd generation quark doublet still has a flat profile). 
As discussed above, IR localization enhances the coupling to graviton KK modes. 
As a result, the 750 GeV resonance has a large and potentially observable branching fraction for decays into a pair of  top quarks  or of Higgs bosons.  
In fact, the parameters of the MAX benchmark are chosen such that the $t \bar t$ and $hh$ rates are close to saturating the experimental limits from the LHC run-1, see \tref{RSBKT_rgamma}.  
Moreover, for this  benchmark, the the first KK graviton is strongly self-coupled. 
Indeed, the cubic self-coupling of the first KK graviton can be estimated as  
\bea
c_{3X} & \approx  & {v \over M_{P}}  { \int_0^L \left ( 1 + r_0 \delta(y) + r_L \delta (y-L) \right ) f_1(y)^3 
\over \int_0^L \left ( 1 + r_0 \delta(y) + r_L \delta (y-L) \right ) f_1(y)^2 f_0 } 
\nnl & \approx  & 
 {v \over a_L M_{P}}  \sqrt{r_0/ r_L}.
 \eea 
 Then, the scale suppressing the cubic coupling in the Lagrangian  is $\Lambda_{3X} \equiv v/c_{3X}$. 
For the parameters of the MAX benchmark  one finds $\Lambda_{3X}  \approx 200$~GeV, which is below the mass of the graviton. 
This leads to a loss of perturbative unitarity in the $X X \to XX$ scattering process shortly above the scale $2 m_X$.  
Further increasing the IR localization of top and Higgs (while maintaining the diphoton signal strength of the order 5~fb) would lead to violating the experimental and perturbativity bounds. 

\begin{table}[h]
\bc 
\begin{tabular}{|c|ccccccc|}
\hline
& IR & MIN & MED & MAX & GMAX && Allowed
\\ \hline \hline
$r_{ZZ}$ & 0.9  & 0.9 & 1.1 & 5.7 & 5.0  && 10~\cite{Aad:2015kna} 
\\ 
$r_{Z\gamma}$ & 0 & 0 & 0 & 0 & 0.5  && 6~\cite{Zgamma}
\\ 
$r_{WW}$ & 1.9 & 1.9 & 2.2 & 10.9 & 9.1   && 40~\cite{Khachatryan:2015cwa}   
\\ 
$r_{hh} $ & 0.06  &  0  & 0.06 & 2.7 & 2.9 &&  40~\cite{ATLAS:2014rxa}
\\ 
$r_{tt}$ & 1.1 &0 &1.2 & 170  & 24  && 400~\cite{Chatrchyan:2013lca}  
\\ 
$r_{e^+e^-+\mu^+\mu^-}$ & 1 & 0 &0 &0 & 0   && 1~\cite{dileptons}   
\\ \hline 
\end{tabular}
\ec 
\caption{
\label{tab:RSBKT_rgamma}
The ratio $r_{ff} \equiv {\Gamma(X \to ff) / \Gamma(X \to \gamma \gamma) }$ for various final states $f$  for the IR,  MIN, MED, MAX, and GMAX  benchmarks  defined in the text. 
The benchmarks are constructed such that this ratio is always smaller than the maximum experimentally allowed one.  
All experimental constraints are taken from~\cite{Franceschini:2015kwy} and rescaled for the updated central value of the di-photon excess, except the bounds on $r_{Z\gamma}$ and $r_{e^+e^-+\mu^+\mu^-}$ which include the recent ATLAS searches at 13TeV LHC~\cite{Zgamma, dileptons}.
}
\end{table}

We have thus shown that one can realize the 750 GeV diphoton resonance as a spin-2 KK graviton in the RS scenario with bulk SM fields. 
However, the result may not be completely satisfactory from the point of view of naturalness. 
Indeed, even for the MAX scenario the Higgs field is only mildly localized towards the IR brane. 
On the other hand, in concrete models addressing Higgs naturalness, a sharper localization is typically predicted. 
For example, if the Higgs doublet arises from a 5th component of the gauge field, 
its  profile corresponds to $\alpha = -1$~\cite{Agashe:2004rs}. 
Such a sharp localization is not possible in the setup discussed so far, because it would leads to an excessive branching fractions of the 750 GeV resonance into $hh$, $ZZ$, and $WW$. 
There is a possible remedy, however. 
Notice that the symmetries of the RS framework allow not only the graviton but also the SM  fields to have brane kinetic terms \cite{Dvali:2000rx,delAguila:2003bh}. 
It turns out that in the presence of large brane kinetic terms for the $SU(2)$ or $U(1)$ SM gauge bosons one can achieve sharp IR localization of the Higgs without violating  experimental bounds. 
For simplicity, in the following we will assume that only the $U(1)$ hypercharge field has a sizable brane kinetic term  on the IR brane, while for the remaining SM fields the brane kinetic terms are negligible. 

The kinetic Lagrangian for the 5D hypercharge field in the warped background of \eref{metric} is parametrized as  
\beq
\cL  \supset  M_*   \left [  - {1 \over 4} B_{\mu \nu} B_{\mu \nu} \left ( 1+  d_L \delta(y-L) \right )
+ {a^2 \over 2} \left (\partial_y B_\mu \right )^2 \right ],  
\eeq  
where the magnitude of the brane kinetic term is controlled by the  parameter $d_{L}$ with dimension $[{\rm mass}]^{-1}$. 
The 5D field is expanded into KK modes  $B_\mu(x,y) =  \sum_n  B_{\mu,n}(x) f_{B,n} (y)$. 
The 5th component of the hypercharge field provides  the Goldstone bosons eaten by massive KK modes, and it is suppressed in this discussion. 
The profile functions satisfy the equations of motion,  
$\partial_y \left ( a^2 \partial_y f_{B,n} \right )  + m_n^2 f_{B,n} = 0$, 
the boundary conditions  
$\partial_y f_{B,n}(0)  = 0$,  $a_L^2  \partial_y f_{B,n}(L)  =  d_L m_n^2 f_{B,n}(L)$, and the orthonormality condition  
$M_* \int_0^L dy  \left (1  +  d_L \delta (y- L) \right ) f_{B,n}(y) f_{B,m}(y ) = \delta_{nm}$. 
The normalized zero mode has a constant profile
$f_{B,0} = {1 / \sqrt{L   + d_L}}$.  
The coupling of the hypercharge  to the first  graviton KK mode is given by  
\bea 
c_{B}  &=&  {v \over M_P} \int_0^L dy \left (1 + d_L \delta (y-L) \right )
{f_1(y) \over f_0} f_{B,0}  
\nnl &\approx & 
 {v  \over   a_L M_P  } \sqrt{r_0 \over r_L}  { 1 + 4 k d_L \over  4 k (d_L  +L)} . 
\eea
The important observation here is that the suppression of this coupling  by $4 k L$ can be countered by brane kinetic terms, provided $d_L k \gtrsim 1$. 
Thus, in the presence of large brane kinetic terms for the gauge fields,   the coupling of KK gravitons to  the zero mode gauge fields  can be comparable to that to the IR localized matter fields. 
In a sense, the IR  brane kinetic terms lead to (quasi)-localization of the gauge field at the IR brane \cite{Dvali:2000rx}.
The net effect is to increase the KK graviton couplings to photons (and Z) relative to other SM fields. 
One nice consequence is that the total production cross section required to fit the diphoton excess is smaller, which allows one to choose smaller 5D parameters for phenomenologically viable benchmarks. 
Thus the model becomes more {\em weakly} coupled as $d_L$ increases;  in particular,  the self-coupling of the KK graviton can be weaker.  
We note however that,  in the context of the 750 GeV spin-2 resonance,    $d_L$ cannot be increased without penalty. 
For large $d_L$, the mass of the first hypercharge KK mode can be approximated by 
\beq
m_1 \approx  2 k  a_L \sqrt {d_L k  +  k L  \over 2 d_L L k^2 + kL -  k d_L  - 1  }\,.
\eeq 
For $d_L \gg L$ one finds  $m_1 \approx \sqrt 2 k a_L/\sqrt{k L}$, and the suppression factor $\sqrt{k L} \sim 5$ may push the mass of the first KK mode dangerously low.

One set of parameters that  fits the ATLAS and CMS diphoton excess without being excluded by resonance searches in other channels  is the following: 
\bea 
\label{eq:GBKT_bench} 
& 
a_L =   10^{-15}, \ 
k = 2.4 \times 10^{18}~\gev, \
\nnl
&M_* = 2.1 \times 10^{17}~\gev,  
\nnl 
&
r_0 = 1500/k, \ r_L = 40/k, \ d_L = 2/k \,, 
\nnl  & 
\alpha_{t_R} = -1,  \ \alpha_{Q_L^3} = 0,  \ \alpha_H = -1. 
& 
\eea 
We refer to this benchmark point as {\bf GMAX}. 
The KK graviton branching fractions are summarized in  the last  column of \tref{RS_br}.
The distinguishing  feature is that  the Higgs and right-handed top fields are sharply localized towards the IR brane. 
In particular, the Higgs profile is the same as predicted by natural models of gauge-Higgs unification.  
Nevertheless, the benchmark is not excluded by run-1 diboson and $t \bar t$ resonance searches, as can be seen in the last column of \tref{RSBKT_rgamma}. 
Also, it is actually {\em less} strongly coupled than the MAX benchmark: the scale suppressing cubic self-interactions of the 1st KK graviton  is $\Lambda_{3X} \approx 400$ GeV.  
These features are thanks to the relative enhancement of the coupling to photons due to the large brane kinetic term $d_L$. 
The third generation quark doublet is assumed to have a flat profile, 
while the remaining fermion fields are localized on the UV brane. 
The KK graviton branching fractions are summarized in  the last  column of \tref{RS_br}.
Note that, unlike for other RS benchmarks discussed in this paper, 
the GKBT benchmark predicts a non-zero branching fraction to $Z \gamma$. 
This is due to non-universal couplings of the KK graviton to  $SU(2) \times U(1)$ SM gauge bosons, which arise as a consequence of the brane kinetic term for the U(1) factor.
The  benchmark  predicts  $c_G = -4.5\times 10^{-3}$, $\sigma(pp \to X)_{13\tev} \approx 0.23$~pb, and  $\sigma(pp \to X \to \gamma \gamma)_{13\tev}  \approx 5.5$~fb, which is again adjustable (for example, by varying $r_0$).

A consequence of the large $d_L$ is  a fairly light KK mode of the hypercharge gauge boson which may be observed by $Z'$ resonance searches at the LHC. 
For the parameters in the GMAX benchmark one finds $m_{B_1} \approx 2.3$~TeV. 
This resonance can be  produced at the LHC in  $q \bar q$  collisions. 
For light quarks localized on the UV brane the couplings to  the gauge KK modes are somewhat suppressed, by a factor $\sim 0.3$, compared to the SM hypercharge,  which results in a suppressed production cross section of $B_1$.  
For the 8 TeV LHC  we find $\sigma(pp \to V_1)_{8\rm TeV} \approx 1$~fb, 
and for  the 13 TeV LHC  we find $\sigma(pp \to V_1)_{13\rm TeV}  \approx 10$~fb. 
This KK mode  decays mostly to fermions localized near the IR brane, while the branching fractions to UV-localized fermions are suppressed. 
We estimate ${\rm Br}(V_1 \to e^+ e^-) = {\rm Br}(V_1 \to \mu^+ \mu^-) \approx 2\%$ and  ${\rm Br}(V_1 \to t \bar t) \approx 87\%$  
For these parameters, a $2.3$~TeV hypercharge KK mode is not excluded by existing dilepton or $t \bar t$ resonance searches~\cite{dileptons, Aad:2015fna}. 

 ~
 
\mysection{Discussion and conclusions}
\label{sec:Conclusion}

We have shown that the 750 GeV diphoton resonance observed by ATLAS and CMS can be realized as the first KK mode of the graviton in the RS scenario. 
The previous realization in Ref.~\cite{Giddings:2016sfr} with all SM fields localized on the IR brane leads to an experimental tension due to non-observation of 750 GeV dilepton resonances in the LHC run-1 and run-2. 
We have shown that this tension can be completely removed when the SM fields propagate in the bulk, and leptons are localized away from the IR brane. 
A consistent realization of this scenario requires introducing large kinetic (Einstein-Hilbert) terms on the UV and IR branes, in order to avoid light KK modes of the SM bulk gauge fields. 
We have argued that there exist large portions of the parameter space that fit the observed mass and cross section of the diphoton resonance without violating experimental bounds in other decay channels, and have presented several concrete benchmark points to demonstrate this. 
Moreover, it is  possible to localize the Higgs field near the IR brane, so that the scenario is compatible with similar models discussed in previous literature which address the hierarchy problem. 
It would be interesting to construct along these lines  a complete model with a 750 GeV spin-2 resonance and a  calculable and natural Higgs potential. 

Phenomenological predictions  of the model depend on the localization of SM matter fields in the extra dimension.
If, as required by naturalness arguments, the zero modes of the Higgs and top fields are localized near the IR brane, one expects large branching fractions  to $t \bar t$, $hh$, $WW$ and $ZZ$.
Decays to $Z \gamma$ can also be observable in the presence of large non-universal brane kinetic terms for the SM $SU(2)\times U(1)$ fields.
Another  prediction of our model is the presence of KK modes of gauge and, possibly (depending on the localization properties), of  other SM fields. 
In the presence of large brane kinetic terms, the lightest KK mode of the SM gauge fields is expected to be within the experimental reach of the LHC. 
As we discussed, the existence of these KK modes with relatively low masses, $m_{KK} \lesssim 5$~TeV, is required on general grounds by perturbative unitarity of the effective theory of a spin-2 particle with non-universal couplings to matter.  
On the other hand, the 2nd KK mode of the graviton is typically very heavy; for our benchmarks with SM in the bulk its mass is $m_{X_2} \sim 6$~TeV which is out of reach of the LHC.  
This is in contrast to the RS model with the SM on the IR brane which predicts that the first spin-2 KK mode at $m_{X_1} \approx 750$~GeV is followed by another at $m_{X_2} \approx 1.4$~TeV.

The holographic interpretation of the RS framework is that it is a dual description of a strongly interacting sector which has an approximate conformal symmetry~\cite{Contino:2003ve}. 
The original RS scenario with the SM on the IR brane corresponds to the case when all SM particles arise as composite states of the strong dynamics. 
On the other, the scenario with bulk SM fields discussed in this paper corresponds to the case where elementary SM particles mix with composite states of the strong sector, where the mixing angle is determined by the localization properties of the SM fields~\cite{Agashe:2004rs}. 
Our effective RS model suggests that there exist strongly interacting theories where a spin-2 state is parametrically the lightest composite resonance.

\begin{acknowledgments}
We thank Kaustubh~Agashe and Riccardo~Rattazzi for the discussion about the radion ghost issue in our construction. 
We also thank Fabio~Maltoni, Kentarou~Mawatari, Fran\c{c}ois~Richard, and Jean-Baptiste~de~Vivie for pointing out errors in the previous version of  \eref{gamma} and \eref{theta}.
J.F.K. would like to thank Gian~Giudice, Matthew~McCullough and Riccardo~Torre for useful discussions, the CERN TH Department for hospitality while this work was being completed and acknowledges the financial support from the Slovenian Research Agency (research core funding No. P1-0035).
AF~is supported by the ERC Advanced Grant Higgs@LHC.
 
\end{acknowledgments}

 ~
 
  ~

\begin{appendix}

\mysection{Appendix: radion}
\label{sec:radion}

In this Appendix we discuss the radion degree of freedom in the RS model with brane kinetic terms for the graviton. 
For simplicity, we set the UV brane kinetic term to zero, $r_0 = 0$, as it plays a less prominent role in radion dynamics.  
We parametrize the fluctuations of the 5D metric as 
\bea
\label{eq:5df_metric}
ds^2 & = & a^2(y)\left ( \eta_{\mu \nu} + h_{\mu\nu}(x,y)  -  {1 \over 2} \eta_{\mu \nu} \phi(x,y)  \right )  dx_\mu   dx_\nu 
\nnl &+ & 2 a^2(y) A_\mu(x,y)  dx_\mu dy  -  \left ( 1 + \phi(x,y) \right ) dy^2 , 
\eea  
where $a = e^{-ky}$. 
The vector fluctuations  described by $A_\mu$ correspond only to non-physical degrees of freedom  eaten by massive gravitons and are ignored in the following.  
With this parametrization, the quadratic Lagrangian for tensor and scalar fluctuations is  given by  
\bea
\label{eq:ads_l2s}
   { \cL_{2} \over M_*^{3}}  & =  & {a^2 \over 4}  \left [ 
{1 \over 2} (\partial_\rho h_{\mu \nu})^2  -  {1 \over 2} (\pa_\rho h)^2 
 -   (\partial_\rho h_{\mu \rho})^2 +   \pa_\mu h   \partial_\rho h_{\mu \rho}   \right ] 
  \nnl & \times  &  \left [ 1 + r_L \delta(y-L) \right ]
-  {a^2 \over 4} \phi \left [ \Box h - \partial_\mu \partial_\nu h_{\mu\nu} \right ] r_L \delta (y-L ) 
   \nnl & + & 
  {3 \over 16} a^2  (\partial_\mu \phi)^2  \left [ 1 - r_L \delta(y-L) \right ]  
 \nnl &-  & 
 {a^4 \over 8 }  \left ( \partial_y h_{\mu \nu}  - {1 \over 2 a^2} \eta_{\mu \nu}   \partial_y  (a^2 \phi)  \right )^2 
 \nnl 
 & +  &  {a^4 \over 8 }    \left ( \partial_y h  - 2  a^{-2} \partial_y  (a^2 \phi)  \right )^2  \,.
\eea
The scalar component, much as the tensor one, can be expanded into KK modes as 
$\phi = \sum_n \tilde f_n(y) \phi_n(x)$. 
Most of the scalar modes are eaten by massive gravitons and do not correspond to new physical degrees of freedom. 
However, there is always the radion mode, here denoted by $\phi_0$, which does not mix with the graviton KK tower and therefore it corresponds to  a physical scalar particle. 
For $r_L = 0$, the canonically normalized radion is described by the profile 
\beq
\label{eq:radion}
\tilde f_0(y) =  a_L a^{-2}(y) \sqrt{2 \over 3} {2  \over  M_P}. 
\eeq
For $r_L > 0$, the radion profile is no-longer described by \eref{radion} because $\phi_0$ will mix with the graviton tower. 
However, the mixing terms have two derivatives, thus integrating out the heavy gravitons produces only four- and higher-derivative radion terms in the low-energy effective Lagrangian \cite{Coradeschi:2013gda}.  
Therefore, to read off the radion kinetic terms, it is sufficient to insert the parameterization of \eref{radion} into the 5D Lagrangian in \eref{ads_l2s}, and integrate over $y$. 
Note that a positive $r_L$ implies  a negative contribution to the radion kinetic term. 
One finds that for 
\beq
k r_L >  1/2 \,,
\eeq
the negative contribution dominates over the positive one from the bulk, and the radion becomes a ghost.  
The same conclusion can be reached with a  proper KK expansion for the tensor and scalar modes in the presence of the IR brane kinetic term,  such that the radion does not mix with the gravitons. 
It remains to be seen whether the problem can be avoided after a stabilization mechanism for the radion is included.

\end{appendix}


\begin{thebibliography}{99}

\bibitem{ATLAS-CONF-2015-081}
  The ATLAS collaboration,
  ATLAS-CONF-2015-081.

\bibitem{CMS:2015dxe}
  CMS Collaboration,
  CMS-PAS-EXO-15-004.

\bibitem{CMS:2016owr} 
  CMS Collaboration [CMS Collaboration],
  CMS-PAS-EXO-16-018.
  
\bibitem{Franceschini:2015kwy} 
  R.~Franceschini {\it et al.},
  arXiv:1512.04933 [hep-ph].
   
\bibitem{Falkowski:2015swt} 
  A.~Falkowski, O.~Slone and T.~Volansky,
  JHEP {\bf 1602}, 152 (2016)
  doi:10.1007/JHEP02(2016)152
  [arXiv:1512.05777 [hep-ph]].
  
\bibitem{Kamenik:2016tuv} 
  J.~F.~Kamenik, B.~R.~Safdi, Y.~Soreq and J.~Zupan,
  arXiv:1603.06566 [hep-ph].
  
  \bibitem{others}
 For a complete list of relevant references see \url{http://inspirehep.net/search?ln=en&p=refersto%3Arecid%3A1410174}
  
\bibitem{Han:2015cty}
  M.~T.~Arun and P.~Saha,
  arXiv:1512.06335 [hep-ph];
  C.~Han, H.~M.~Lee, M.~Park and V.~Sanz,
  Phys.\ Lett.\ B {\bf 755}, 371 (2016)
  [arXiv:1512.06376 [hep-ph]];
J.~S.~Kim, K.~Rolbiecki and R.~R.~de Austri,
 arXiv:1512.06797 [hep-ph].
 M.~R.~Buckley,
 arXiv:1601.04751 [hep-ph];
  A.~Martini, K.~Mawatari and D.~Sengupta,
  arXiv:1601.05729 [hep-ph];
C.~Q.~Geng and D.~Huang,
  arXiv:1601.07385 [hep-ph];
  V.~Sanz,
  arXiv:1603.05574 [hep-ph].

\bibitem{Giddings:2016sfr} 
  S.~B.~Giddings and H.~Zhang,
  arXiv:1602.02793 [hep-ph].
 
   
  
    \bibitem{Dvali:2000hr} 
  G.~R.~Dvali, G.~Gabadadze and M.~Porrati,
  Phys.\ Lett.\ B {\bf 485}, 208 (2000)
  [hep-th/0005016].

\bibitem{Han:1998sg} 
  T.~Han, J.~D.~Lykken and R.~J.~Zhang,
  Phys.\ Rev.\ D {\bf 59}, 105006 (1999)
  doi:10.1103/PhysRevD.59.105006
  [hep-ph/9811350].
 
\bibitem{Lee:2013bua} 
  H.~M.~Lee, M.~Park and V.~Sanz,
  Eur.\ Phys.\ J.\ C {\bf 74}, 2715 (2014)
  doi:10.1140/epjc/s10052-014-2715-8
  [arXiv:1306.4107 [hep-ph]].
  
\bibitem{Mathews:2005bw} 
  P.~Mathews, V.~Ravindran and K.~Sridhar,
  JHEP {\bf 0510}, 031 (2005)
  [hep-ph/0506158].
  

\bibitem{Gao:2010bb} 
  J.~Gao, C.~S.~Li, B.~H.~Li, C.-P.~Yuan and H.~X.~Zhu,
  Phys.\ Rev.\ D {\bf 82}, 014020 (2010)
  [arXiv:1004.0876 [hep-ph]].
  
\bibitem{Ball:2014uwa} 
  R.~D.~Ball {\it et al.} [NNPDF Collaboration],
  JHEP {\bf 1504}, 040 (2015)
  [arXiv:1410.8849 [hep-ph]].

\bibitem{Artoisenet:2013puc} 
  P.~Artoisenet {\it et al.},
  JHEP {\bf 1311}, 043 (2013)
  [arXiv:1306.6464 [hep-ph]].


\bibitem{Randall:1999ee} 
  L.~Randall and R.~Sundrum,
  Phys.\ Rev.\ Lett.\  {\bf 83}, 3370 (1999)
  [hep-ph/9905221].

\bibitem{Gibbons:1976ue} 
  G.~W.~Gibbons and S.~W.~Hawking,
  Phys.\ Rev.\ D {\bf 15}, 2752 (1977).
  doi:10.1103/PhysRevD.15.2752

\bibitem{Goldberger:1999uk} 
  W.~D.~Goldberger and M.~B.~Wise,
  Phys.\ Rev.\ Lett.\  {\bf 83}, 4922 (1999)
  [hep-ph/9907447].


\bibitem{Csaki:2000zn} 
  C.~Csaki, M.~L.~Graesser and G.~D.~Kribs,
  Phys.\ Rev.\ D {\bf 63}, 065002 (2001)
  doi:10.1103/PhysRevD.63.065002
  [hep-th/0008151].


\bibitem{Luty:2003vm} 
  M.~A.~Luty, M.~Porrati and R.~Rattazzi,
  JHEP {\bf 0309}, 029 (2003)
  doi:10.1088/1126-6708/2003/09/029
  [hep-th/0303116].

\bibitem{George:2011sw} 
  D.~P.~George and K.~L.~McDonald,
  Phys.\ Rev.\ D {\bf 84}, 064007 (2011)
  doi:10.1103/PhysRevD.84.064007
  [arXiv:1107.0755 [hep-ph]].


\bibitem{Aad:2014cka} 
  G.~Aad {\it et al.} [ATLAS Collaboration],
  Phys.\ Rev.\ D {\bf 90}, no. 5, 052005 (2014)
  [arXiv:1405.4123 [hep-ex]].

\bibitem{dileptons} 
  The ATLAS collaboration,
  ATLAS-CONF-2015-070.


\bibitem{Davoudiasl:1999tf} 
  H.~Davoudiasl, J.~L.~Hewett and T.~G.~Rizzo,
  Phys.\ Lett.\ B {\bf 473}, 43 (2000)
  [hep-ph/9911262].

\bibitem{Pomarol:1999ad} 
  A.~Pomarol,
  Phys.\ Lett.\ B {\bf 486}, 153 (2000)
  [hep-ph/9911294].

\bibitem{Bajc:1999mh} 
  B.~Bajc and G.~Gabadadze,
  Phys.\ Lett.\ B {\bf 474}, 282 (2000)
  [hep-th/9912232].

\bibitem{Chang:1999nh} 
  S.~Chang, J.~Hisano, H.~Nakano, N.~Okada and M.~Yamaguchi,
  Phys.\ Rev.\ D {\bf 62}, 084025 (2000)
  [hep-ph/9912498].


\bibitem{Grossman:1999ra} 
  Y.~Grossman and M.~Neubert,
  Phys.\ Lett.\ B {\bf 474}, 361 (2000)
  [hep-ph/9912408]; 
  T.~Gherghetta and A.~Pomarol,
  Nucl.\ Phys.\ B {\bf 586}, 141 (2000)
  [hep-ph/0003129]; 
  S.~J.~Huber and Q.~Shafi,
  Phys.\ Lett.\ B {\bf 498}, 256 (2001)
  [hep-ph/0010195];
  K.~Agashe, G.~Perez and A.~Soni,
  Phys.\ Rev.\ D {\bf 71}, 016002 (2005)
  [hep-ph/0408134].

\bibitem{Manton:1979kb} 
  N.~S.~Manton,
  Nucl.\ Phys.\ B {\bf 158}, 141 (1979);
  Y.~Hosotani,
  Phys.\ Lett.\ B {\bf 126}, 309 (1983); 
  H.~Hatanaka, T.~Inami and C.~S.~Lim,
  Mod.\ Phys.\ Lett.\ A {\bf 13}, 2601 (1998)
  [hep-th/9805067];
  I.~Antoniadis, K.~Benakli and M.~Quiros,
  New J.\ Phys.\  {\bf 3}, 20 (2001)
  [hep-th/0108005].
  
\bibitem{Agashe:2004rs} 
  K.~Agashe, R.~Contino and A.~Pomarol,
  Nucl.\ Phys.\ B {\bf 719}, 165 (2005)
  [hep-ph/0412089].
  


  
  
  
\bibitem{Kiritsis:2002ca} 
  E.~Kiritsis, N.~Tetradis and T.~N.~Tomaras,
  JHEP {\bf 0203}, 019 (2002)
  [hep-th/0202037].

\bibitem{Davoudiasl:2003zt} 
  H.~Davoudiasl, J.~L.~Hewett and T.~G.~Rizzo,
  JHEP {\bf 0308}, 034 (2003)
  [hep-ph/0305086].

\bibitem{Aad:2015kna} 
  G.~Aad {\it et al.} [ATLAS Collaboration],
  Eur.\ Phys.\ J.\ C {\bf 76}, no. 1, 45 (2016)
  [arXiv:1507.05930 [hep-ex]].

\bibitem{Khachatryan:2015cwa} 
  V.~Khachatryan {\it et al.} [CMS Collaboration],
  JHEP {\bf 1510}, 144 (2015)
  [arXiv:1504.00936 [hep-ex]].
  
\bibitem{ATLAS:2014rxa} 
  The ATLAS collaboration [ATLAS Collaboration],
  ATLAS-CONF-2014-005.

\bibitem{Zgamma} 
  The ATLAS collaboration,
  ATLAS-CONF-2016-010.

\bibitem{Chatrchyan:2013lca} 
  S.~Chatrchyan {\it et al.} [CMS Collaboration],
  Phys.\ Rev.\ Lett.\  {\bf 111}, no. 21, 211804 (2013)
  Erratum: [Phys.\ Rev.\ Lett.\  {\bf 112}, no. 11, 119903 (2014)]
  doi:10.1103/PhysRevLett.111.211804, 10.1103/PhysRevLett.112.119903
  [arXiv:1309.2030 [hep-ex]].


\bibitem{Dvali:2000rx} 
  G.~R.~Dvali, G.~Gabadadze and M.~A.~Shifman,
  Phys.\ Lett.\ B {\bf 497}, 271 (2001)
  [hep-th/0010071].

\bibitem{delAguila:2003bh} 
  F.~del Aguila, M.~Perez-Victoria and J.~Santiago,
  JHEP {\bf 0302}, 051 (2003)
  [hep-th/0302023].

\bibitem{Aad:2015fna} 
  G.~Aad {\it et al.} [ATLAS Collaboration],
  JHEP {\bf 1508}, 148 (2015)
  [arXiv:1505.07018 [hep-ex]].
  
\bibitem{Contino:2003ve}
 R.~Contino, Y.~Nomura and A.~Pomarol,
 Nucl.\ Phys.\ B {\bf 671}, 148 (2003)
 [hep-ph/0306259].
 
\bibitem{Coradeschi:2013gda} 
  F.~Coradeschi, P.~Lodone, D.~Pappadopulo, R.~Rattazzi and L.~Vitale,
  JHEP {\bf 1311}, 057 (2013)
  doi:10.1007/JHEP11(2013)057
  [arXiv:1306.4601 [hep-th]].
  
  
\end{thebibliography}
\end{document}